# HYLU: Hybrid Parallel Sparse LU Factorization

Xiaoming Chen
Institute of Computing Technology, Chinese Academy of Sciences, Beijing, China
chenxiaoming@ict.ac.cn

***Abstract*----**This article introduces HYLU, a hybrid parallel LU factorization-based general-purpose solver designed for efficiently solving sparse linear systems (**Ax**=**b**) on multi-core shared-memory architectures. The key technical feature of HYLU is the integration of hybrid numerical kernels so that it can adapt to various sparsity patterns of coefficient matrices. Tests on 37 sparse matrices from SuiteSparse Matrix Collection reveal that HYLU outperforms Intel MKL PARDISO in the numerical factorization phase by geometric means of 2.36X (for one-time solving) and 2.90X (for repeated solving). HYLU can be downloaded from https://github.com/chenxm1986/hylu.

## 1. Introduction

Solving sparse linear systems is a core task in scientific computing. Two types of methods to solve sparse linear systems are direct and iterative methods. Direct methods are generally suitable for small- to medium-scale sparse linear systems. Large-scale sparse linear systems are mostly solved by preconditioned iterative methods, as the time and memory overheads of direct methods will become unbearable.

The high performance of traditional sparse direct linear solvers heavily rely on gathering structurized nonzero elements into dense blocks and computing them using the level-3 basic linear algebra subroutines (BLAS), especially the general matrix multiplication (GEMM) function. Popular implementations include supernode- and multifrontal-based algorithms, based on which people have developed some software packages, such as UMFPACK [1], SuperLU [2], PARDISO [3], etc.

However, for a big portion of practical sparse linear systems which are highly sparse (e.g., linear systems from circuit simulation and power network simulation), such level-3 BLAS-based methods are inefficient. Instead, dedicated solvers have been developed for circuit simulation problems, including KLU [4], NICSLU [5], CKTSO [6], etc. They generally perform well on circuit matrices but may not efficiently solve linear systems from other areas. How to design a general-purpose sparse direct solver that can efficiently solve linear systems with a variety of sparsity patterns is a challenge.

This article introduces HYLU, a hybrid parallel LU factorization-based general-purpose solver designed for efficiently solving sparse linear systems (**Ax**=**b**) on multi-core shared-memory architectures. To efficiently solve linear systems with various sparsities, HYLU integrates multiple numerical kernels and incorporates a smart kernel selection strategy based on the matrix sparsity, such that it can adapt to various sparse linear systems. Tests on 37 sparse matrices from SuiteSparse Matrix Collection [7] reveal that HYLU outperforms Intel MKL PARDISO in the numerical factorization phase by geometric means of 2.36X (for one-time solving) and 2.90X (for repeated solving).

## 2. Brief Technical Description of HYLU

Completely solving a sparse linear system by LU factorization needs 3 main phases: preprocessing, numerical factorization, and forward-backward substitution. This section briefly introduces the 3 steps of HYLU.

### 2.1 Preprocessing

The preprocessing phase of HYLU includes 3 steps: static pivoting, ordering for fill-in reduction, and symbolic factorization. HYLU adopts the maximum weighted matching algorithm [8] for static pivoting. It permutes large elements to the diagonal, and also scales the matrix values, such that the diagonal elements are 1 or -1, and non-diagonal elements are bounded in [-1,1]. After static pivoting, matrix reordering is performed to minimize fill-ins which will be generated during LU factorization. The approximate minimum degree (AMD) algorithm [9], a modified implementation of AMD [10], and a modified nested dissection algorithm based on METIS [11], are adopted in HYLU for reordering. Finally a symbolic factorization is performed to form the symbolic structure of the LU factors, which will not change during numerical factorization. The number of floating-point operations is calculated during symbolic factorization, and supernodes are also detected. HYLU will select the numerical kernel based on these numbers and other information.

## 2.2 Numerical Factorization

The sparse left-looking algorithm [12] is widely adopted by sparse linear solvers. HYLU uses the row-major order by default, so it uses the transposed version of the left-looking algorithm, which can be called an *up-looking* algorithm. The ordinary left-looking algorithm, which is adopted by KLU, is only suitable for extremely sparse matrices, such as matrices from circuit simulation problems. The concept of *supernode* [2] is widely used to explore the regularity in irregular sparse matrices of LU factorization. By gathering nonzero elements into dense sub-blocks, BLAS functions can be employed to accelerate sparse LU factorization. In HYLU, a supernode is defined as a set of consecutive rows which have the identical structure in **U**.

To make HYLU a general-purpose sparse linear solver that can efficiently solve a wide range of sparse linear systems, HYLU integrates multiple up-looking numerical kernels: a) a row-row kernel, b) a sup-row kernel, and c) a sup-sup kernel, as illustrated in Fig. 1. The row-row kernel is the ordinary up-looking algorithm. In this kernel, no BLAS function is called. The numerical computation is implemented by plain C code. The sup-row kernel still updates a row at a time, but uses supernodes as source data. In this case, level-2 BLAS can be called to accelerate sup-row updates. In the sup-sup kernel, each time HYLU uses a supernode to update a supernode. Level-3 BLAS is employed to accelerate sup-sup updates. After all updates from previous supernodes and rows are completed, an internal factorization is performed to factorize the self supernode, which also employs level-3 BLAS functions. In the sup-row and sup-sup kernels, there exist many standalone rows that do not belong to any supernode. For such rows, using level-3 BLAS functions is an overkill and has no performance benefit. Instead, HYLU uses the row-row or sup-row kernel to perform such row-related numerical updates. The data structure of HYLU is elaborated to support the hybrid numerical kernels in a common way.

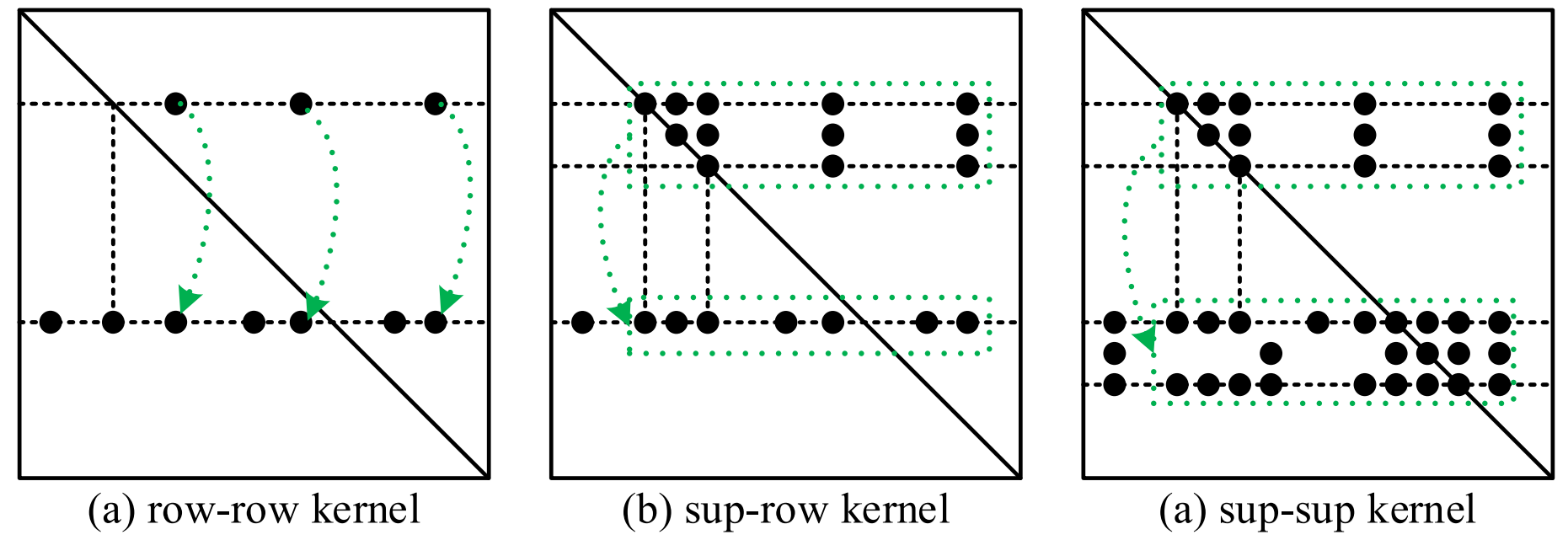

(a) row-row kernel (b) sup-row kernel (a) sup-sup kernel

Fig. 1: Three numerical kernels integrated in HYLU.

Supernode diagonal pivoting is performed during numerical factorization. For each row in a supernode, the maximum element in its diagonal block is exchanged to the diagonal. For standalone rows, pivoting is not performed as there is no other choice. Pivot perturbation [13] is also performed to replace small or zero pivots.

### 2.2.1 Parallel Numerical Factorization

The parallelism of numerical factorization of HYLU is realized by analyzing the inter-row/-supernode dependency and utilizing a dual-mode parallel factorization scheme [14], which is similar to the parallelism of NICSLU and CKTSO. Fig. 2 illustrates the parallelization methodology. The symbolic structure of **L** implies the data dependency. A task dependency graph can be drawn from symbolic factorization, where each node is a standalone row or a supernode. The dependency graph can be levelized such that nodes in the same level are independent. A dual-mode parallelism methodology can be applied. Front levels have many nodes in each level, so each level can be computed in parallel, and a barrier is needed after each level. This is called a *bulk mode*. The remaining levels have much stronger dependency as they usually form a long dependent chain. A *pipeline mode* is proposed to explore finer-grained parallelism between dependent nodes.

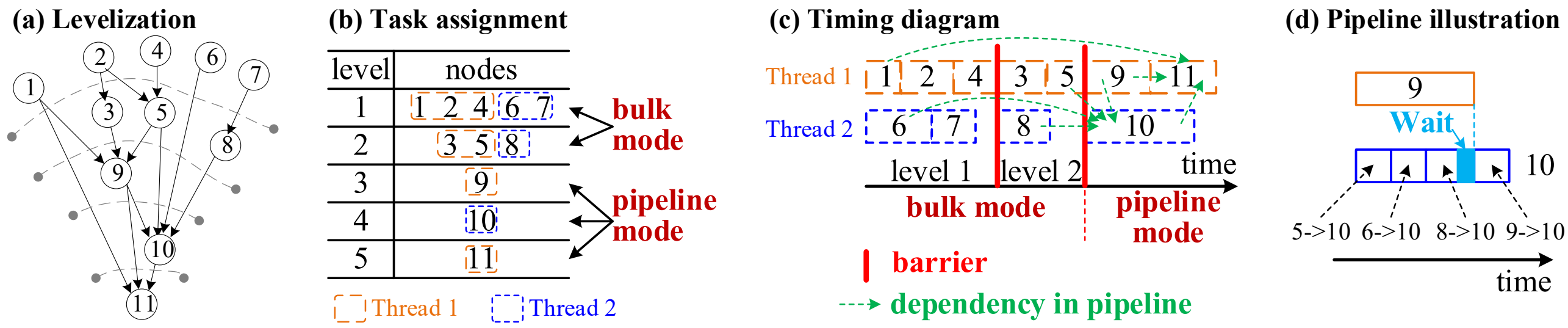

Fig. 2: Levelization-based dual-mode parallel scheduling method [14].

### 2.3 Forward-Backward Substitution

The forward-backward substitution phase is also parallelized. HYLU adopts a matrix partition-based parallel solving method similar to that of CKTSO. Fig. 3 illustrates the partitioning method. For the largest triangular part, a dual-mode solving method is used. Different from parallel numerical factorization, here the dual-mode means "bulk-sequential", which means that levels with many nodes on each level are solved by the bulk mode and the long chain is solved sequentially. Each rectangular block is solved in parallel and different rectangular blocks are processed in sequential, while the workload balance is achieved by evenly assigning nonzeros in each rectangular block to threads. The small triangular blocks are solved sequentially.

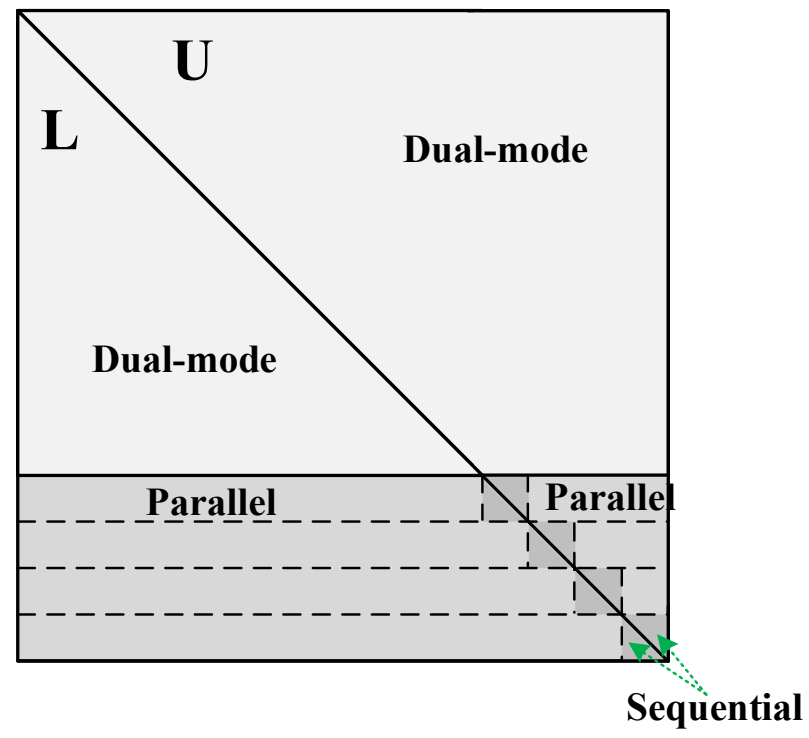

Fig. 3: Parallel forward-backward substitution by matrix partitioning.

If pivot perturbation has occurred during numerical factorization, iterative refinement [13] will be automatically executed during solving.

## 3. Test Results

The experiments were carried out on a Linux server. The main hardware and software configurations are listed in Table I. All results presented in this section are wall-time measurements from 16-thread parallel execution.

Table I: Hardware and software configurations of performance tests.

| | |
|---|---|
| CPU | Intel Xeon Gold 6130 @ 2.1GHz |
| Memory | 256GB |
| Operating system | Ubuntu 24.04 LTS for tests/CentOS 7.9 for compiling HYLU |
| Compiler | gcc 13.3.0 for compiling test code/gcc 4.8.5 for compiling HYLU |
| MKL version | 2025.3 |
| HYLU version | 20260331 |
| Benchmarks | 37 matrices from SuiteSparse Matrix Collection, dimensions from 525,825 to 5,558,326 |

### 3.1 One-Time Solve

On geometric mean, HYLU achieves a **2.36X** speedup in numerical factorization compared with Intel MKL

PARDISO, while the preprocessing phase is **1.48X** faster. The forward-backward substitution phase of HYLU is slightly slower and is 18% slower on geometric mean than that of Intel MKL PARDISO. For the total one-time solve time (preprocessing + numerical factorization + forward-backward substitution), HYLU is **1.70X** faster than Intel MKL PARDISO on geometric mean.

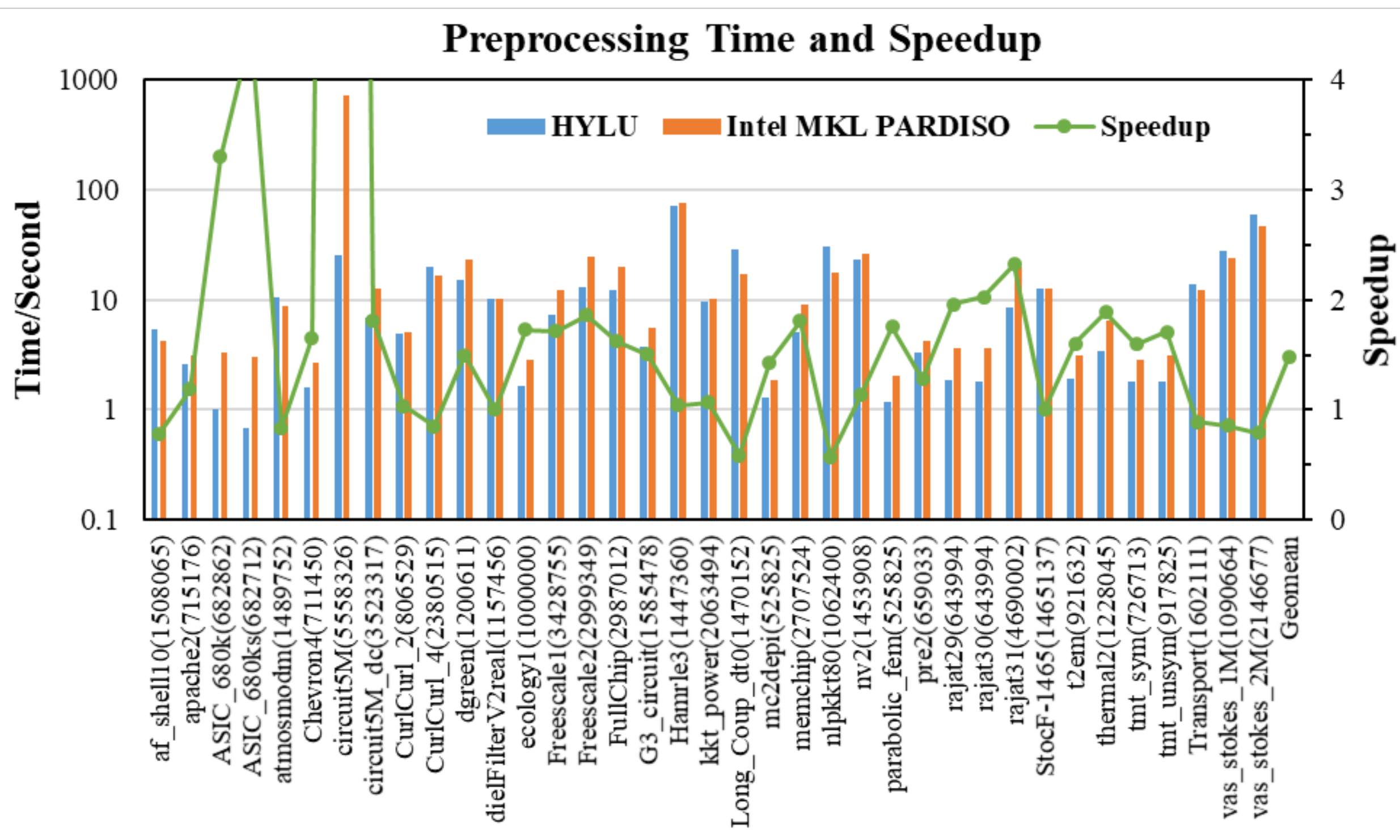

Fig. 4: Preprocessing time and speedup for one-time solving.

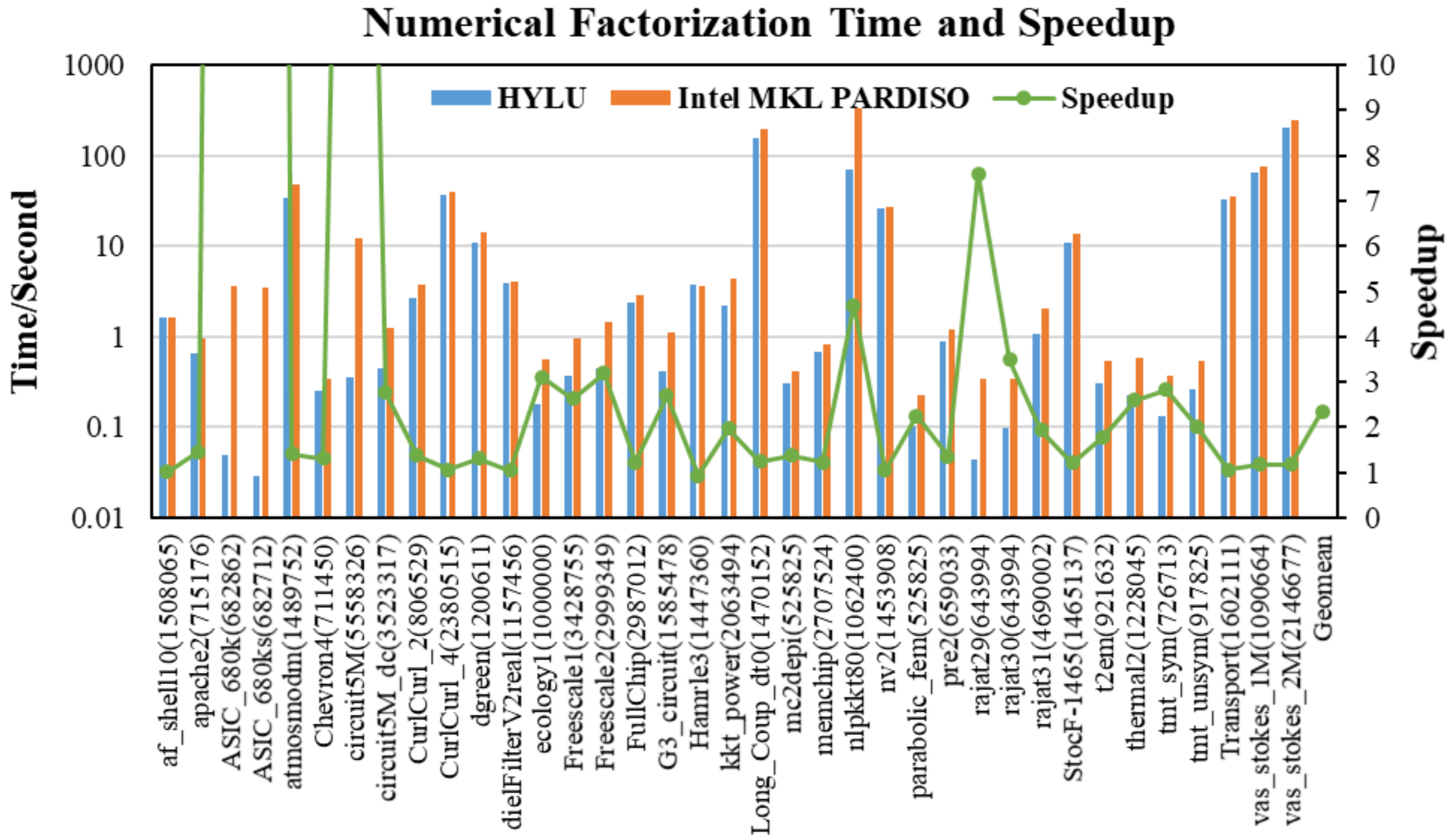

Fig. 5: Numerical factorization time and speedup for one-time solving.

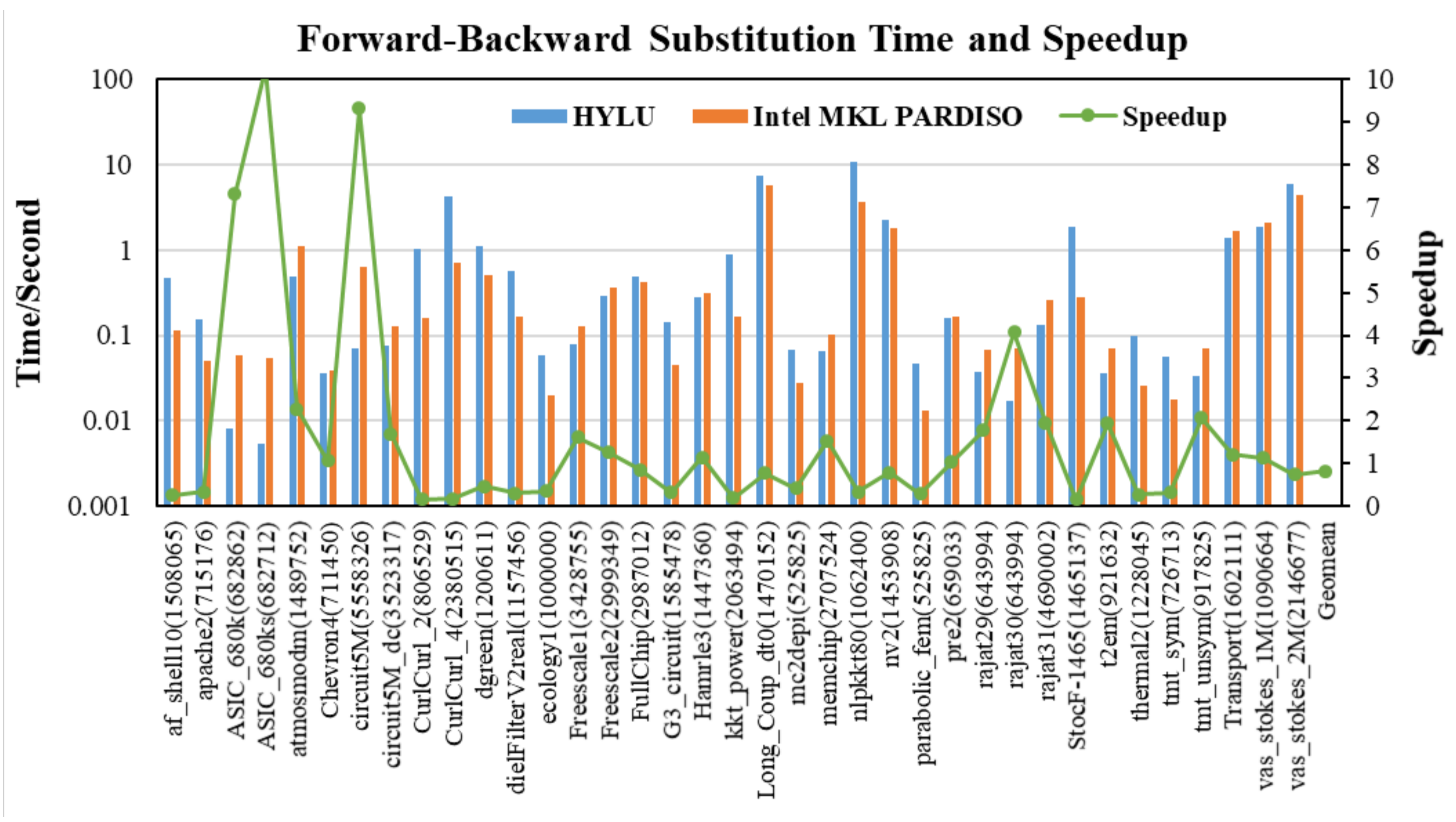

Fig. 6: Forward-backward substitution time and speedup for one-time solving.

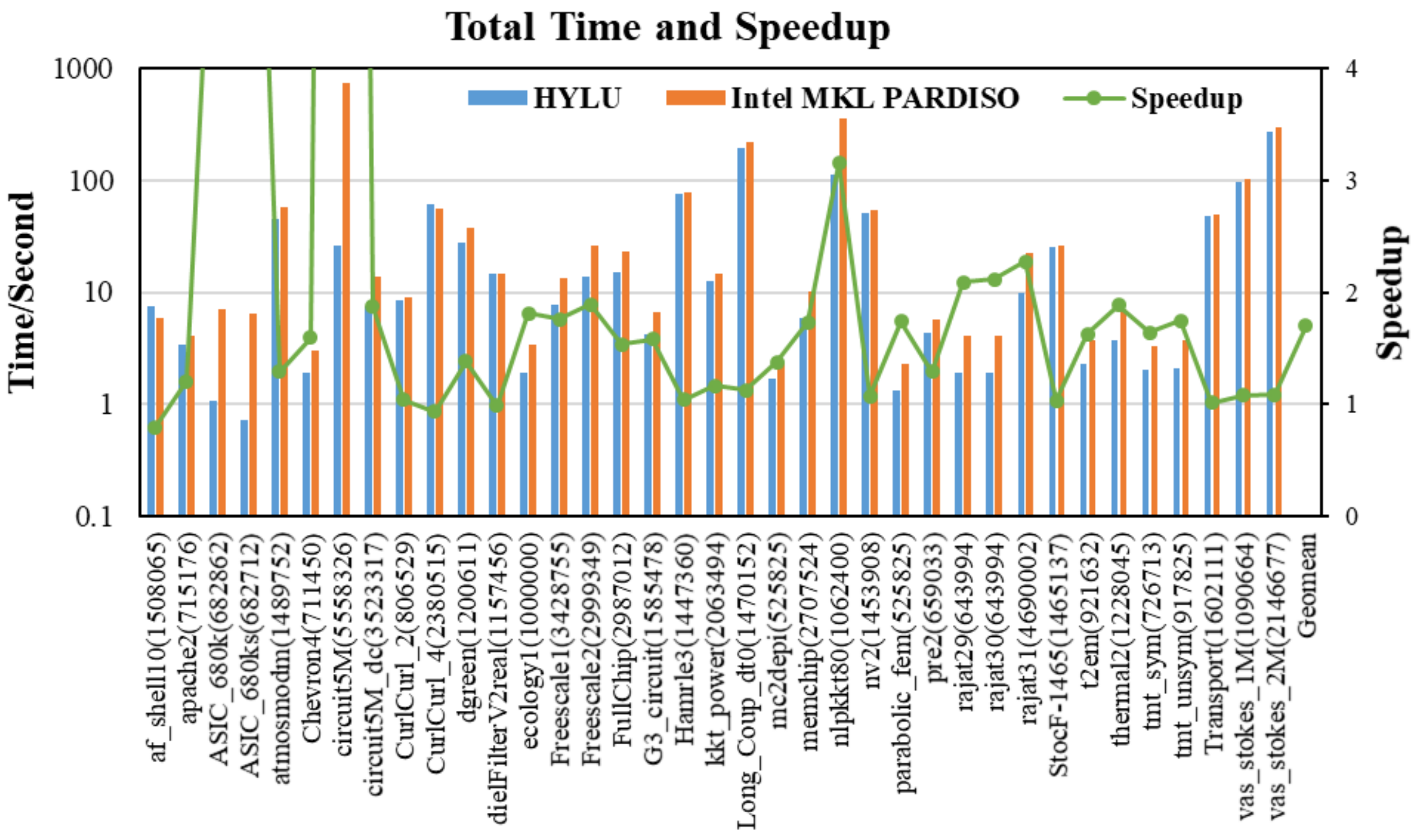

Fig. 7: Total time and speedup for one-time solving.

Another advantage of HYLU is its stable performance for various sparsities. Intel MKL PARDISO generates a huge quantity of fill-ins for some benchmarks (ASIC_680k, ASIC_680ks, circuit5M, and nlpkkt80), and the performance of these matrices is poor.

### 3.2 Repeated Solve

HYLU offers an optimization option for repeated solve of linear systems with an identical sparse pattern in the coefficient matrix. In this case, HYLU achieves a **2.90X** geometric mean speedup in numerical factorization over Intel MKL PARDISO, while the forward-backward substitution phase is slightly slower (20% slower on geometric mean). When comparing the total time of numerical factorization and forward-backward substitution, HYLU is faster than Intel MKL PARDISO for **ALL** of the tested benchmarks and the geometric mean of speedups is **2.53X**. In the repeated solve scenario, the preprocessing phase is on geometric mean 1.75X slower than that of the one-time solve scenario. However, in the repeated solve scenario, the pre-processing time is less important, as it is executed only once.

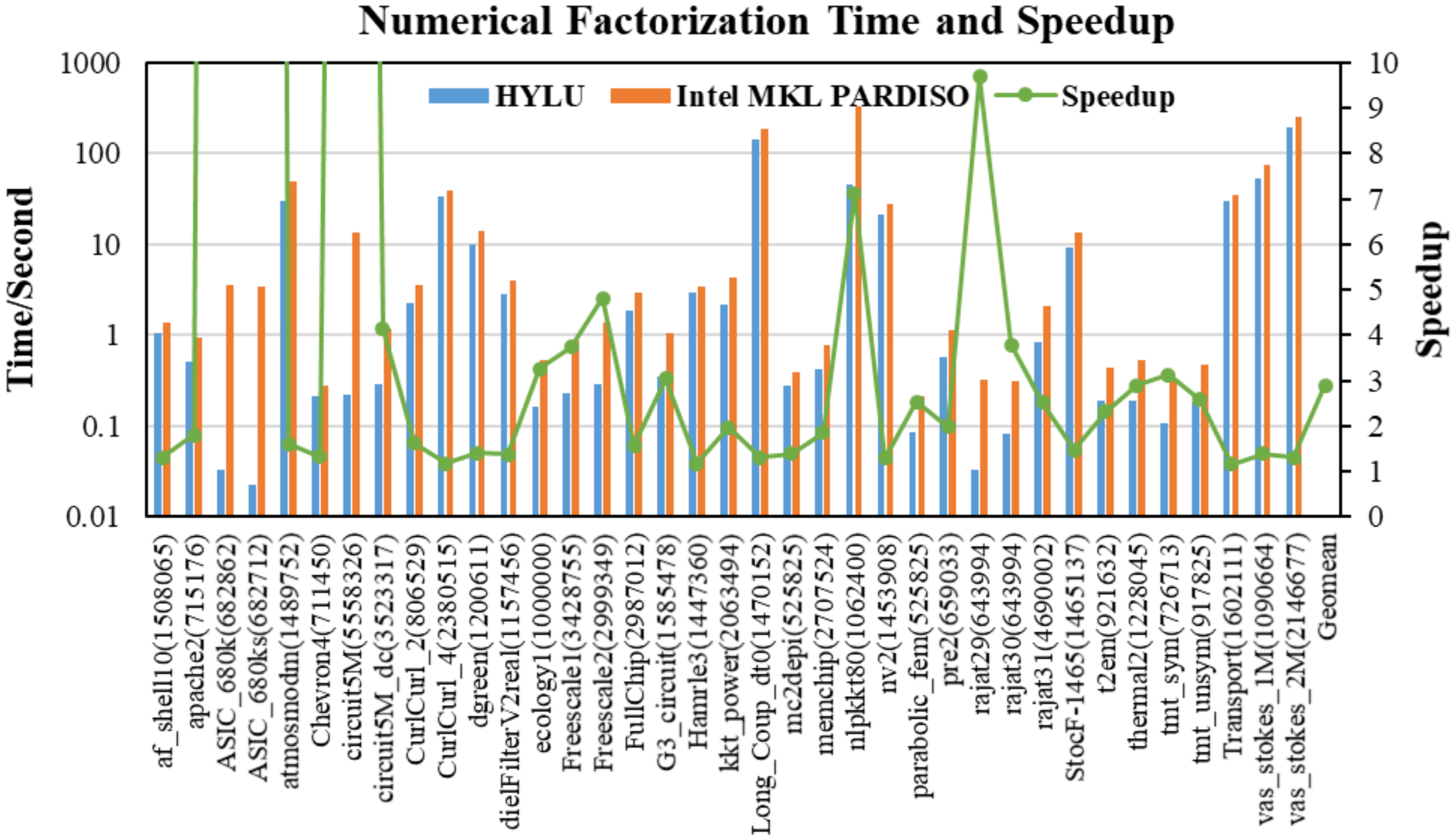

Fig. 8: Numerical factorization time and speedup for repeated solving.

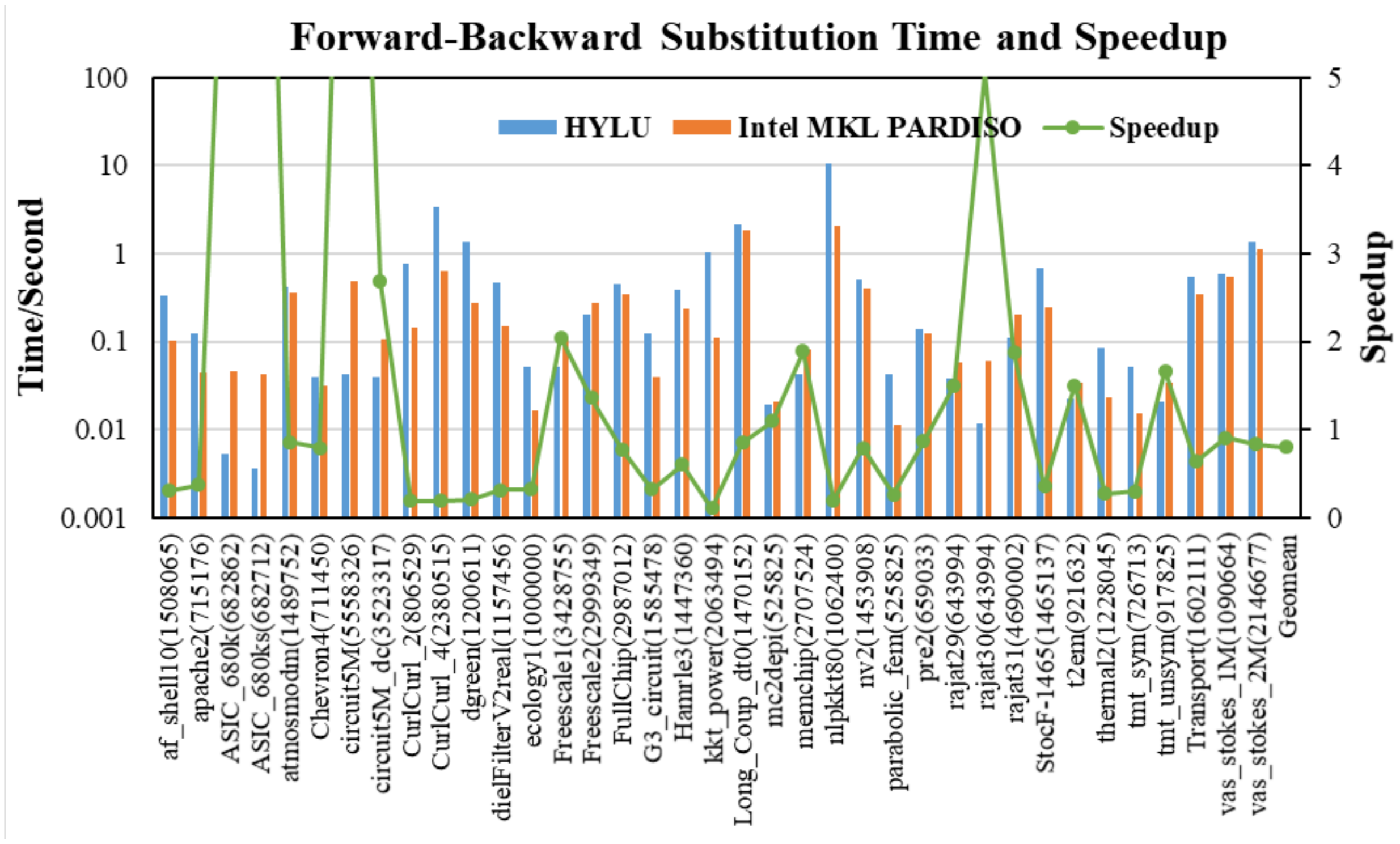

Fig. 9: Forward-backward substitution time and speedup for repeated solving.

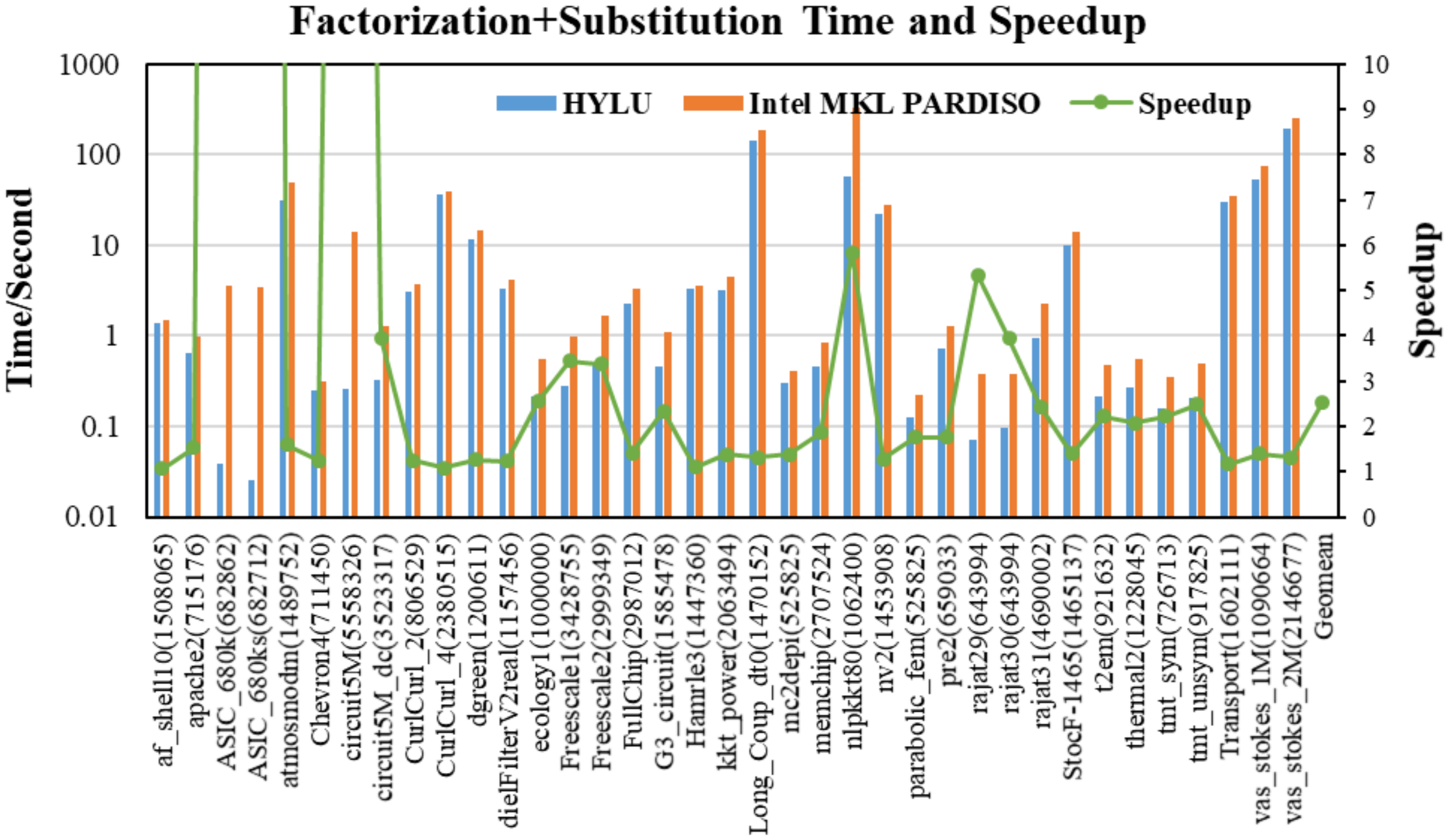

Fig. 10: Total time of factorization and substitution and speedup for repeated solving.

### 3.3 Accuracy

The residual is calculated as $\frac{\|\boldsymbol{A}\boldsymbol{x}-\boldsymbol{b}\|_1}{\|\boldsymbol{b}\|_1}$. On geometric mean, HYLU achieves **an order of magnitude higher accuracy** than Intel MKL PARDISO. This is mainly due to better control of pivoting and iterative refinement of HYLU, where the latter, however, introduces some overhead to the forward-backward substitution phase. Both HYLU and Intel MKL PARDISO fail to accurately solve Hamrle3, due to its extremely large condition number.

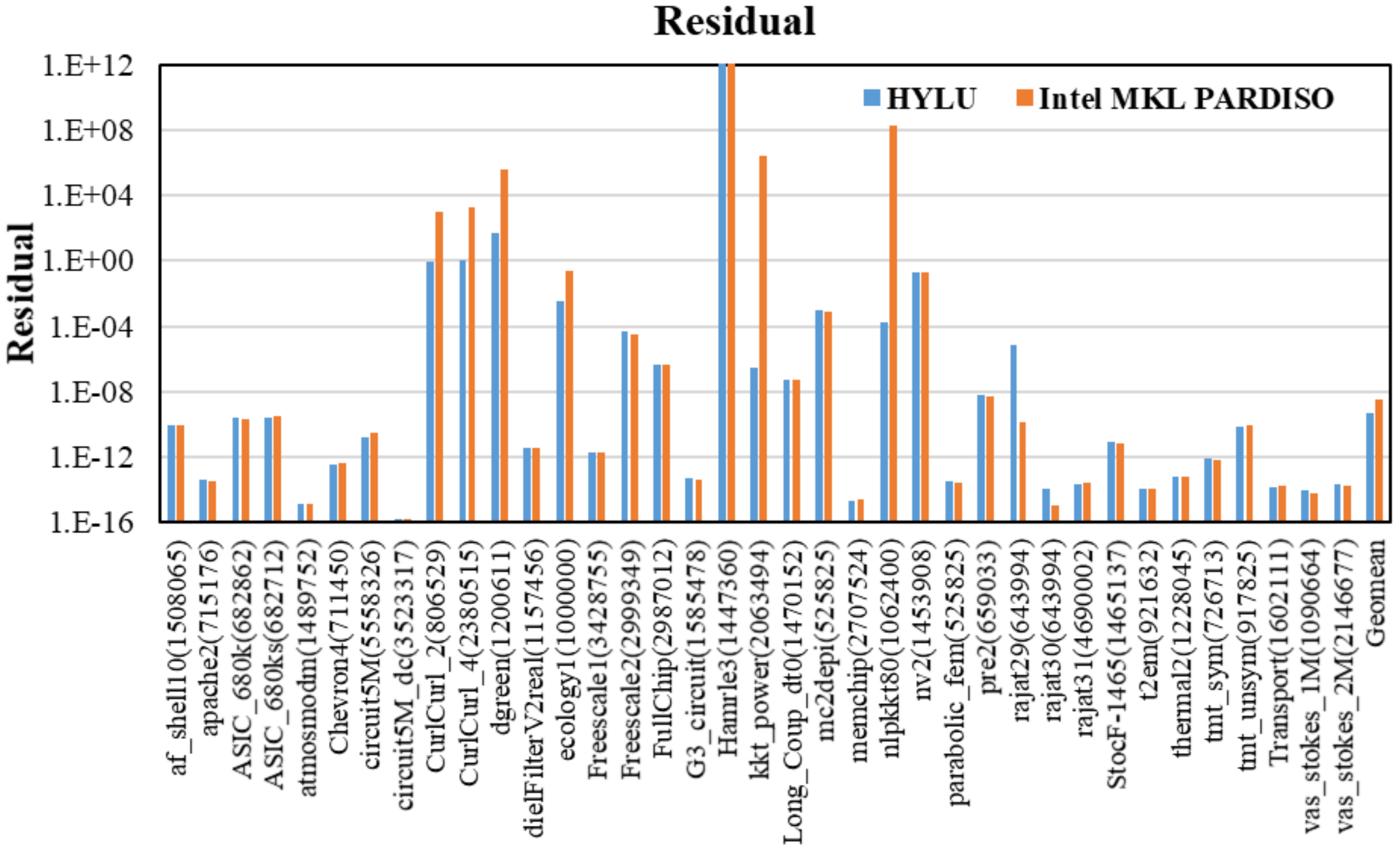

Fig. 11: Residual.

## 4. Conclusions

Though the idea of supernodes has been adopted by many popular sparse direct solvers, it is not always an efficient method for various sparsities of linear systems. Instead, integrating different numerical kernels and elaborately selecting them based on the matrix sparsity pattern can be more efficient for a wide range of sparsities. Based on this idea, HYLU is developed, which shows more superior performance in numerical factorization compared with Intel MKL PARDISO.